\documentstyle[12pt,epsf,axodraw]{article}
\begin{document}

\newcommand{\be}{\begin{equation}}
\newcommand{\ee}{\end{equation}}
\newcommand{\bea}{\begin{eqnarray}}
\newcommand{\eea}{\end{eqnarray}}
\newcommand{\lbl}[1]{\label{eq:#1}}
\newcommand{\ceq}[1]{(\ref{eq:#1})}
\newcommand{\cfig}[1]{\ref{#1}}
\newcommand{\matr}{\left( \begin{array}}
\newcommand{\ematr}{\end{array} \right)}
\newcommand{\dis}{\displaystyle}
\newcommand{\ba}{\begin{array}}
\newcommand{\ea}{\end{array}}
\newcommand{\beqa}{\begin{eqnarray}}
\newcommand{\eeqa}{\end{eqnarray}}

\newcommand{\lsim}{{\;\raise0.3ex\hbox{$<$\kern-0.75em\raise-1.1ex
\hbox{$\sim$}}\;}}
\newcommand{\gsim}{{\;\raise0.3ex\hbox{$>$\kern-0.75em\raise-1.1ex
\hbox{$\sim$}}\;}}

\thispagestyle{empty}

   \begin{flushright}
   HIP-1998-71/TH \\
   hep-ph/98xxxx
   \end{flushright}

\vspace*{3cm}

\begin{center}
{\bf \large
Production of $q\:q\:\bar q\:\bar q$ final states in $e^-e^-$ collisions in the left-right symmetric model 
}
\end{center}

\vspace*{3cm}

\centerline{J. Maalampi$^a$ and N. Romanenko$^{a,b}$}

\begin{center}
{\footnotesize \it
 $^a$ Theoretical Physics 
  Division, Department of Physics,
   University of Helsinki, Finland \\
 $^b$ Petersburg Nuclear Physics Institute,
   Gatchina, Russia}
\end{center}

\vfill
\abstract{We consider the reaction $e^-e^- \rightarrow 
q\:q\:\bar q\:\bar q$ as
 a test of lepton number non-conservation 
 in the framework of the left-right-symmetric
  electroweak model. The main contributions to this process
   are due to  Majorana neutrino exchange in $t$-channel 
   and  doubly charged Higgs ($\Delta^{--}$) exchange in $s$-channel
with a pair of right-handed weak bosons ($W_R$)
 as intermediate state. We show that
in a linear $e^-e^-$ collider with the collision energy of  1 TeV (1.5 TeV)
the cross section of this process is 0.01 fb (1 fb),
 and it will, for the anticipated
  luminosity of $10^{35}$  cm$^{-2}$,
   be detectable below the $W_R$ threshold. We 
   study the sensitivity of the reaction on the masses 
   of the heavy neutrino, $W_R$ and $\Delta^{--}$.}

\vspace*{15mm}
\vfill

\newpage

\section{Introduction}
\setcounter{equation}{0}
  The electroweak model with the left-right (LR) gauge symmetry
$SU(3)_c \otimes SU(2)_L \otimes SU(2)_R \otimes U(1)_{B-L}$,  
proposed in \cite{LR}, is one of the most popular extensions 
of the Standard Model (SM). It gives a better understanding of parity
violation than SM and it maintains the lepton-quark symmetry  in weak interactions. 
Parity is in it  broken spontaneously, and embedding of the model into the SO(10) grand unified scheme \cite{Ell} can be implemented consistently when the
scale of the discrete LR-symmetry breaking is  more than 1 TeV or so.

Perhaps the most important property of the LR-model is its ability to provide, in terms of the seesaw mechanism
\cite{SS}, a simple and  natural explanation to  the smallness of  the masses of the ordinary neutrinos.
This results from the mixing of the ordinary left-handed neutrinos 
with right-handed neutrinos, which quite naturally achieve 
a Majorana mass of the order of 2--3
 $W_R$-masses  \cite{AR}. The ordinary
  neutrinos are predicted by this model
   to be very light, but -- in contrast with the
    SM -- not exactly massless, Majorana particles.
 The recent observation by the SuperKamiokande experiment of the atmospheric neutrino oscillations  \cite{Kam} confirmed that
 at least some of the neutrino species do have a mass, giving an
 additional argument in favour of the LR-symmetric model.

An essential ingredient of the LR-model are the triplet  scalars. They are needed to break the LR-symmetry in a consistent way so that at low energies the model  reproduces
the SM interactions and at the same time give rise to the seesaw mass mechanism of neutrinos.  Their interactions with fermions break the lepton number by two units, $\vert\Delta L\vert=2$, as do the Majorana mass terms of neutrinos they give rise to. The $e^-e^-$--col\-lisions give the most pure environment to study the $\vert\Delta L\vert=2$ interactions, because the corresponding
 SM background is suppressed as the lepton number is conserved in the SM. In the literature  different observable lepton number violating processes, including
 doubly charged Higgs production \cite{MaalD2},
 vector-boson pair and triple production for
 elect\-ron-po\-sitron and electro-electron colliders 
  \cite{MaalW2,Gun}, have been investigated.

 In the present paper we will study the 
   lepton-number violating process

\be
e^-e^-\to q\:q\:\bar q\:\bar q
\label{reaction}\ee
with various quark flavour combinations.
This process, as it breaks the lepton number, 
is forbidden in the  SM.
 One would expect to obtain indirect
  evidence of the LR-model via this process well 
  below the threshold of $W_R^{\pm}$, 
  the gauge boson of the right-handed interactions,
   and other new particles predicted by the  model.
  
According to the existing plans the
 Next Linear Collider (NLC) will
operate at energies up to $\sqrt{s}\approx 1-2$
TeV, and it is assumed to have a
 luminosity of the order of $10^{35}$  cm$^{-2}$
   \cite{coll}.   We will show  in this paper that 
   with this kind of equipment
    it will be possible  to detect the  reaction (\ref{reaction})
for a reasonable choice of the parameters of the
LR-symmetric model and obtain quite strong mass constraints for the new gauge and Higgs bosons of the model.  

  The organization or this article is as follows:
  in Section 2 we give the description of particle
  content, lagrangian and general properties of the
  LR-model; in Section 3 we derive the amplitudes of the reaction (\ref{reaction}) and discuss the corresponding reactions with a leptonic final state; 
 in Section 4 we present the
 numerical results of our calculations;
 in Section 5
we discuss the SM background; and
 Section 6 is devoted to conclusions.

\section{Description of the model}  
 \setcounter{equation}{0}

 In the
LR-model quarks and leptons are assigned to the following $SU(2)_L\times SU(2)_R\times U(1)_{B-L}\,$
representations
\cite{Rabibook}:
\be 
{ Q}_{iL}=
\left[ \begin{array}{c} u \nonumber \\ d
\nonumber \end{array} \right]_{iL}=({2}, 1,
\frac{1}{3});\: 
{ Q}_{iR}=
\left[ \begin{array}{c} u \nonumber \\ d
\nonumber \end{array} \right]_{iR}=(1, {2},
\frac{1}{3}) \ee 
\be 
{ \Psi}_{iL}=
\left[ \begin{array}{c} \nu \nonumber \\ e
\nonumber \end{array} \right]_{iL}=({2}, 1, -1);\:
{ \Psi}_{iR}=
\left[ \begin{array}{c}
\nu \nonumber \\
e \nonumber \end{array} \right]_{iR}=(1, {2}, -1),
\ee
where $i$
is the flavour index.
 In addition to the SM particles, each family 
contains a right-handed neutrino. The gauge sector differs from the SM due to presence of
right-handed gauge bosons $W_R$ and $Z_R$. 
The scalar sector should contain essentially more
particles than in the SM.
In order to generate fermion masses one  needs
the Higgs bidoublet with the following quantum numbers:
$$\Phi=\left(
\begin{array}{ll}
\phi_1^0 & \phi_1^+ \nonumber \\
\phi_2^- & \phi_2^0
\end{array}
\right)=({2}, {2}^*,0),
$$
and with the following vacuum expectation value (VEV)
$$ <\Phi>= \frac{1}{\sqrt2}\left(
\begin{array}{ll}
k_1 & 0 \\
0   &k_2
\end{array} \right)
$$
This is, however, not enough to accomplish the spontaneous symmetry breaking of the 
gauge group $SU(2)_L\times SU(2)_R\times U(1)_{B-L}\,$
into the SM symmetry, but some other Higgs field with non-vanishing $B-L$
is needed.
 There are several alternatives for 
 the additional Higgs multiplet
 \cite{MDL}, but if one wants to generate neutrino  masses 
 through the seesaw mechanism, the triplet
 Higgs field  $\Delta_R$, sometimes also called a Higgs-Majoron, is needed:

\be
 {\dis\Delta_R=\matr{cc}\Delta_R^+/ \sqrt{2}&\Delta_R^{++}\\
\Delta_R^0&-\Delta_R^+ /\sqrt2\ematr = (1,3,2)}
\ee
with the vacuum expectation value:
\be
\begin{array}{c}  {\dis\langle\Delta_{R}\rangle
=\frac1{\sqrt{2}}\matr{cc}0&0\\v_{R}&0
\ematr.}
\end{array}
\ee  
If one imposes an explicit $L \leftrightarrow R$
symmetry, the corresponding left-handed
Higgs-Majoron field should also be introduced:
\be
 {\dis\Delta_L=\matr{cc}\Delta_L^+/ \sqrt{2}&\Delta_L^{++}\\
\Delta_L^0&-\Delta_L^+ /\sqrt2\ematr = (3,1,2)}
\ee
with the vacuum expectation value:
\be
\begin{array}{c}  {\dis\langle\Delta_{L}\rangle
=\frac1{\sqrt{2}}\matr{cc}0&0\\v_{L}&0
\ematr.}
\end{array}
\ee  
As far as masses of neutrinos and gauge bosons are concerned,
 the presence of the 
 left-handed Higgs-Majoron is not, however, essential.

The most general potential describing self-interactions
  of the scalar fields introduced above can be found, e.g.,
  in \cite{MDL}. There exist severe 
  phenomenological bounds on the parameters
  of this potential, particularly  from the limitations on the flavour changing neutral
  currents (FCNC). Since the choice
  $v_L=k_2=0$ satisfies these bounds \cite{MDL} we will
  restrict ourselves in this paper to this case.
  This choice means in particular that we do not allow any mixing
  between charged vector boson fields
  $W_L$ and $W_R$. Then the masses of the charge 
vector bosons are given by the expression
  
  \bea {M_{W_L}^2}&=&\frac{1}{4}g_L^2k_1^2,\\
{M_{W_R}^2}&=&\frac{1}{4}g_R^2(2v_R^2+k_1^2).
\eea

In the case of explicit left-right symmetry
gauge couplings of both 
$SU(2)$ groups should be equal
($g_R=g_L\simeq 0.64$). Without this symmetry
the internal consistency within  the model requires
nevertheless
$g_R \ge 0.55\, g_L$ \cite{Cvetic}. 
The experimental value of the left-handed charged boson
mass is $M_{W_L}=81$ GeV, while the  lower
bound from Tevatron is $M_{W_R}> 650$ GeV \cite{FNAL}. 

 As for the fermion masses, they come from the Yukawa 
 interactions of quarks and leptons:
  $$
-{\cal L }_{Yuk}=
  \bar\Psi_L^i(f_{ij}\Phi +
g_{ij}\tilde\Phi)\Psi_R^j + h.c.+\bar Q_L^i(f^q_{ij}\Phi +
g^q_{ij}\tilde\Phi)Q_R^j + h.c.+
$$
\be
 +h_{R,ij}\Psi_{iR}^TC\sigma_2\Delta_R\Psi_{jR} + 
h_{L,ij}\Psi_{iL}^TC\sigma_2\Delta_L\Psi_{jL}
\:\:\:+\:{\rm h.c.}, 
\lbl{Yuk}
\ee
  where
$\tilde\Phi=\sigma_2\Phi^*\sigma_2$ and $i,j$ are flavour indices.
This yields the usual  quark $3 \times 3$ mass matrix
and charged lepton masses, while for the neutrino
one obtains the seesaw mass matrix:
\be
   M = \left(   \begin{array}{cc} m_L & m_D \\ m_D^T & m_R  \end{array} 
\right).
\label{seesaw}\ee
 The entries are $3\times 3$ matrices given by $m_D=(fk_1 +
gk_2)/\sqrt{2}$,
$m_L = \sqrt2 h_L v_L$ and
$m_R=\sqrt2 h_R v_R$. We will also ignore possible
mixing between the lepton families, so that these matrices are assumed diagonal.
 Natural seesaw condition implies $m_{Di} \approx
 m_{li}$, where $m_{li}$ is the charged lepton mass,
 while the evident phenomenological 
 left-right hierarchy implies $v_R>>k_1$ and hence
 $m_{Ri}>>m_{Di}$. The ensuing  neutrino masses are
$m_{\nu_{1i}}\simeq m_{Di}^2/m_{Ri}$ 
and $m_{\nu_{2i}} \simeq m_{Ri}$.
 The mixing
angle $\eta$ between left-handed and right-handed
neutrino states is given by
\be
   \tan 2\eta_i = \frac{2m_{Di}}{m_{Ri}}.
\label{eta}
\ee 
 Since it is natural that scale of right-handed
 neutrino masses is of order 1--3 $M_{W_R}$
 \cite{AR,MaalW2}, the following values
 of the mixing angle $\eta$
are reasonable:
$$\eta_1 \approx \frac{m_e}{m_R}=0.5 \cdot 10^{-6},$$
$$\eta_2 \approx \frac{m_{\mu}}{m_R}= 10^{-4},$$
$$\eta_3 \approx \frac{m_{\tau}}{m_R}=2 \cdot 10^{-3}.$$
 We will use these values in the following calculations.

\section{ 
Feynman amplitudes }
 \setcounter{equation}{0}

Let us first give the arguments that make us to consider the reaction (\ref{reaction}) particularly suitable for testing  the LR-model. First of all, the final state particles are all light, so that there is no kinematical suppression for the process, in contrast with, e.g., the $W_R$ pair production. Consequently, one may expect 
to detect evidence of the LR-model through this reaction well below the  $W_R$ threshold.
The same is true, of course, for the leptonic final states, for example for the reaction
$e^-e^- \rightarrow \mu^-\:\mu^-\:\mu^-\:\mu^+$. Reactions with ordinary neutrinos in the final state are not very useful as invisibility of neutrinos makes them not easy to distinguish from the background processes. Also, reactions with final state electrons are not 
that good because of the possible mix-up of the initial and final state particles.

  In Fig 1. and Fig. 2 we show Feynman diagrams for the reactions 
$e^-e^- \rightarrow\mu^-\:\mu^-\:\mu^-\:\mu^+$  and
 $e^-e^- \rightarrow
b\:b\:\bar{t}\:\bar{t} $, respectively. The reason for our studying the four-quark final states instead of the four-lepton final states becomes evident from these figures.  One can see that
the reaction with leptons in the final state does not involve
charged vector bosons as intermediate states but  is 
quite sensitive
to the structure of the neutral current sector, while
 the  reaction with quarks in the final state
involves charged vector bosons, particularly the right-handed boson $W_R$, but not the 
neutral ones. There is a variety of extensions
 of the Standard Model where one has extra 
 neutral gauge boson(s),
  such as the superstring-inspired E(6) models \cite{E6}, but no new charged gauge bosons, in contrast with the LR-model.  Hence the  reactions like $e^-e^- \rightarrow
b\:b\:\bar{t}\:\bar{t} $ that  involve charged currents but not neutral currents offer a more unambiguous test  of the LR-symmetric model than the leptonic processes. 

Consequently, we have chosen the processes $e^-e^- \rightarrow
q\:q\:\bar q\:\bar q$ for our further
 investigation. We prefer final states with
  $b$-quarks as the $b$-jets are relatively 
  easy to identify in experiment 
  (the same should be expected for $t$-jets)
    \cite{redbook}.
 From this point of view,
 the best  process for a study would be $e^-e^- \rightarrow
b\:b\:\bar{t}\:\bar{t} $. However, as  will be seen from our numerical results,
it will possible to measure the cross section also  for the  4-jet reactions with no
$b$-jets, as well as  for the reactions with a single $b$-jet.

 The Feynmann graphs for the process  $e^-e^- \rightarrow
b\:b\:\bar{t}\:\bar{t}$ are presented 
in Fig.2.   Some of these
diagrams may be safely neglected without
 any substantial effect on our numerical results.
  First of all, since the
left-handed electron neutrino
 is very light ($m_{\nu_{1}}<1$ eV)
compared with the right-handed one (for which the seesaw mechanism in its simplest form implies
$m_{\nu_{2}} \simeq 1-2$ TeV)
 and  with the collision energy, the amplitudes 2, 3, 6
  and 7 in Fig.2  have, in comparison with, say, diagram 9,
 an extra overall factor of
  $m_{\nu_{1}}/m_{\nu_{2}}$ or
$m_{\nu_{1}}/\sqrt{s}$ due to the lepton-number-violating
neutrino propagator and they may be therefore ignored.
 Diagrams 4, 5 and 8 contain,
  due to neutrino mixing, a small
   parameter  $\sin \eta_1$ in the
     $e\nu_{2}W^+_L$ vertex:
\be
{\cal L}_{e \nu_{2} W_L}
\simeq 
\sin \eta_1 \cdot \frac{g_L}{\sqrt2}  W^+_{L\mu}
\bar{\Psi}_{eL} \gamma_{\mu} \Psi_{\nu_{2L}}
+h.c.,
\ee
and also their contribution can be neglected. Hence there are only two amplitudes, corresponding to the diagrams 1 and 9, that are relevant.

The following lagrangian vertices give rise
to the  diagrams 1 and 9:
\be
 h_{R,11} \cdot \Delta^{--}
 \Psi_{eL}^TC \Psi_{eL}+h.c.,
\ee
where $h_{R,11}$ is defined in \ceq{Yuk}, 
\be
- \frac{g_R^2}{\sqrt2}v_R
\cdot \Delta^{--}W_R^-W_R^- + h.c.
\ee 
which originates in the kinetic term of the Higgs-Majoron
field, and
\be
 -\frac{g_R}{\sqrt2}
 \cdot V_{tb} W^+_{\mu R} \bar{t}_R \gamma_{\mu} b_R.
\ee
 The total amplitude is then:
\newcommand{\gm}{\gamma^{\mu}}
\newcommand{\gr}{\gamma_R}
\newcommand{\gmp}{\gamma^{\mu '}}
\newcommand{\gn}{\gamma^{\nu}}
\newcommand{\gnp}{\gamma^{\nu '}}
\newcommand{\gf}{\gamma_5}

{
$$
M_{ss'rr'qq'}= 
 e_s^T(p_1) T_{\mu \mu '} e_{s'}(p_2) \cdot
\left( i \frac{g_R}{\sqrt2} V_{tb}\right)^2 
\cdot
$$
\be
\frac{1}{\sqrt2}
\left[
\bar{t}_r \gamma^{\mu} \gr b_q \cdot
\bar{t}_{r'} \gamma^{\mu '} \gr b_{q '} 
+
\bar{t}_r \gamma^{\mu} \gr b_{q '} \cdot
\bar{t}_{r'} \gamma^{\mu '} \gr b_{q } 
\right]
\ee}
Here $e_s, b_q, \bar{t}_r$ denote the electron,
 $b$ and $\bar{t}$  4-spinors with the
 corresponding spin indices,
 $\gr \equiv (1+\gf)/2$, $V_{tb}$ is the
 element of the right-handed Kobayasi-Maskawa
 matrix. $T_{\mu \mu '}$ contains the contributions
 from different channels:
 \be
 T_{\mu \mu '} =T^s_{\mu \mu '} +T^t_{\mu \mu '} 
 +T^u_{\mu \mu '}. 
 \ee 
  The $s,t,u$ indices correspond to the Mandelstam
  variables if one treats charged bosons involved
  in the considered Feynman diagramms as a final state
  particles (in other words, each of the two 
  $b,\bar{t}$ clusters is treated as a single particle).
  Then, in correspondence with \cite{MaalW2}, we have:
  {
\be
  T^s_{\mu \mu '} =
  h_{R11}(1+\gf)\cdot 
  \frac{iC \cdot i}{k^2-M^2_{\Delta_{--}}}
  (-2i)\frac{g_R^2}{\sqrt2}
  \cdot g^{\nu \nu '} \Pi_{\mu \nu}(k_1)
  \Pi_{\mu ' \nu '}(k_2)
  \ee 
    \be
  T^t_{\mu \mu '} = \left(\frac{ig_R}{\sqrt2}
  \right)^2 \gn \gr
  \cdot i C^{-1} \frac{\slash{ \hspace*{-2.9mm}p}+m}{p^2-m^2}
  \gnp \gr \Pi_{\mu \nu}(k_1)
  \Pi_{\mu ' \nu '}(k_2)
  \ee
  \be
  T^u_{\mu \mu '}= \left(\frac{ig_R}{\sqrt2} 
  \right)^2 \gn \gr
  \cdot i C^{-1} \frac{\slash{ \hspace*{-2.9mm}p}+m}{p^2-m^2}
  \gnp \gr \Pi_{\mu \nu}(k_2)
  \Pi_{\mu ' \nu '}(k_1) 
  \ee
}
  Here $k$ is the 4-momentum of the doubly charged Higgs, $p$ is the 4-momentum of the Majorana neutrino,
   $k_{1,2}$ are the 4-momentum of the charged
  bosons, and
  {
$$
   \Pi_{\rho \lambda}(q)=\frac{-i}{q^2-M_{W_R}^2 + \frac i2
   \Gamma_{W_R} M_{W_R}}
   \left( g_{\rho \lambda}-
   \frac{q_{\rho} q_{\lambda} }{M_{W_R}^2} \right)    
       $$}
  is the Breit-Wigner  form of massive vector boson 
  propagator ( this is the form
   form the propagators are used
  by  CompHEP \cite{CompHEP}) .
  For the "t-channel" amplitude one has 
  $p=p_1-k_1$, $p+p_2=k_2$, while for the
  "u-channel" amplitude the momentum conservation
  law implies $p=p_1-k_2$, $p+p_2=k_1$.

 We estimate the width of the right-handed
boson  to be 
$\Gamma_{W_R} \approx  \Gamma_{W_L} \cdot M_{W_R}/M_{W_L} $,
and that of the doubly-charged Higgs-Majoron 
$\Gamma_{\Delta^{--}} \approx 0.053\: M_{\Delta^{--}}$.
For the right-handed neutrino we assume
$\Gamma_{\nu_R} \approx 7-70$ GeV for $m_{\nu_2} \sim $
1--2 TeV, but the width has not much effect on our results  since
the right-handed neutrinos are far  from their
pole in our case.

 \section{Numerical results}

 By means of CompHEP \cite{CompHEP} we have derived  the squared matrix elements for $e^-e^- \rightarrow b\:b\:\bar{t}\:\bar{t} $ and computed the ensuing the cross sections  at the collision energies
$\sqrt s=1$ TeV and $\sqrt s=1.5$ TeV. The results depend on a number of unknown parameters of the LR-model, the most important ones
being the masses of the right-handed boson $W_R$ and doubly charged Higgs-Majoron $\Delta^{--}$.
As was discussed before, we 
consider theory without $W_L-W_R$ mixing and neglect
small effects of the seesaw
mixing. We  restrict ourselves
to the manifestly left-right symmetric case, implying that the left and right-handed interactions have the same coupling strength, i.e. $g_L=g_R$,  and
that the  Kobayasi-Maskawa mixings of the right-handed charged currents
are exactly the same as those of the left-handed ones, in particular $V^R_{tb}= V^L_{tb}\equiv V_{tb}$.

We evaluate the values of the gauge coupling constants 
at the linear collider energy scale $\sqrt s$
through one-loop massless renormalization group
equations of the SM without the Higgs boson
contribution:
\begin{eqnarray}
g^2(s)&=&
g^2(M_Z^2)
\left(1+\frac{g^2(M_Z^2)}{16\pi^2}\,
\frac{10}{3}\log\frac{s}{M_Z^2}
\right)^{-1},
\nonumber \\
g'{}^2(s)&=&
g'{}^2(M_Z^2)
\left(1-\frac{g'{}^2(M_Z^2)}{16\pi^2}\,
\frac{20}{3}\log\frac{s}{M_Z^2}
\right)^{-1},
\label{Eq:gs}
\end{eqnarray}
which are related to $e$ and $\sin\theta_W$ as:
\begin{equation}
e^2=\frac{g^2g'{}^2}{g^2+g'{}^2}, \qquad
\sin^2\theta_W=\frac{g'{}^2}{g^2+g'{}^2}.
\end{equation}
Here $M_Z$ is the neutral $Z$-boson mass.
We do not take into account in these equations the additional
particles of the LR-model, making the assumption that they are effectively decoupled due to their large mass.
The effects of any possible  light Higgs particle
are also neglected since they are anyhow relatively small.
 At the energy scale of order of 
 the SM neutral $Z$ boson mass
 ($\sqrt s =M_Z= 91$ GeV) we use the
standard electroweak input \cite{Sch,PDG,Blondel}.

In Fig. \cfig{fe} we show the energy dependence
of the total cross section of the process $e^-e^- \rightarrow
b\:b\:\bar{t}\:\bar{t} $  for various values of
masses of the triplet Higgs $\Delta^{--}$ and the 
right-handed neutrino $\nu_2$. In all the cases
the right-handed boson mass is taken to be $M_{W_R}= 700$ GeV.
We remind that the present experimental
lower bound  from the Tevatron measurements 
is $M_{W_R}\simeq 650$ GeV \cite{FNAL}.

The  plot  Fig. \cfig{fe}a presents the cross section  
for the case of the right-handed neutrino mass 
$m_{\nu_2}=1$ TeV and for three different values
of $M_{\Delta^{--}}$
(600,  1000, 1500 GeV) as indicated in the figure.
The characteristic behaviour of  these curves is that
 they all have a resonance
at $M_{\Delta^{--}}=\sqrt s$ and the  cross section grows by several orders
of magnitude above the $W_R$ threshold.
The  value of the cross section at the resonance 
is determined by the  $\Delta^{--}$ 
width and depends on the assumption
 we made on it in the previous section, while
the growth above the $W_R$ threshold
is easy to understand since
the phase space above the threshold
contains the poles of the
  charged right-handed boson propagator.

 In the plot in Fig. \cfig{fe}b presents the
  cross section in the case $m_{\nu_2}=1.5$ TeV
for the triplet Higgs mass values 
$M_{\Delta^{--}}= 400, \: 800,  \: 1200, \: 2000$
GeV. A comparison with Fig. \cfig{fe}a 
shows that  the increase in the right-handed
neutrino mass makes
 the cross
section larger. The reason for this is obvious:
 the amplitude represented by the diagrams 1 and 9
  of Fig. 2 
 increases with the growth of Yukawa 
 coupling of the neutrino $h_{R,11}$.
 
   As we keep $M_{W_R}$
and $g_R$ fixed, also the  the VEV $v_R$
of the Higgs triplet is fixed, and so
the increase of the right-handed neutrino mass
is solely due to a corresponding increase 
of the Yukawa coupling $h_{R,11}$.
This makes the contribution of the diagrams 1 and 9
larger  than in the case of Fig. \cfig{fe}a.
On the whole, the intimate
 relation of the neutrino mass and 
 lepton number violating couplings 
 in the LR model is directly reflected 
 in the behaviour of the cross sections.


 

In Fig. \cfig{fe}c 
we present the cross section for $m_{\nu_2}=2$ TeV
for three different values of
the triplet Higgs mass, $ M_{\Delta^{--}}=800, \: 1200,
\: 1600, \: 2000 $ GeV. 
 In all cases $M_{W_R}=  700$ GeV.
The value
$m_{\nu_2}=2$ TeV, when $M_{W_R}  =700$ GeV,
corresponds to a value of the coupling 
constant $h_{R,11}$ close to unity.
Going beyond this to higher neutrino masses would not yield reliable results due 
to the unitarity bound.



Given the cross sections, it is interesting to
study what will be the sensitivity of the 
NLC in testing the central
parameters of the LR-model
 through the reaction (\ref{reaction}). In the following we will
present the contours in the $M_{W_R}-M_{\Delta^{--}}$
plane corresponding to various event rates of  $e^-e^- \rightarrow
b\:b\:\bar{t}\:\bar{t} $. We consider the collision
energies 
1 TeV and 1.5 TeV and the anticipated luminosity of $10^{35}$ cm$^{-2} \cdot
{\rm  s}^{-1} $ appropriately scaled 
with the collision energy
(we consider the luminosity to be 
approximately proportional to the collision energy). 
At the collision energy $\sqrt s= 1.5$ TeV
the process with
the cross section $\sigma = 0.01,\: 0.1, \:1$
fb would produce 30, 300 and 3000 events per year,
correspondingly.

In deriving the contours one cannot use the computed cross sections as such but
has to impose several phase space cuts to make quark jets unambiguously identified.
Following the arguments of \cite{redbook}
we apply the following cuts:

\begin{itemize}
\item[--] Each b-jet should have energy more than 10 GeV.
\item[--] Each t-jet should have 
energy more than 190 GeV.
\item[--] The opening 
angle between two detected jets should be
greater than $20^{\circ}$.
\item[--]The angle between each detected jet and
the colliding axis should be greater than 
$36^{\circ}.$
\item[--]The total energy of the event should be
greater than 400 GeV.
\end{itemize}
 
  
In Fig. \cfig{s10n15}
 we display the $M_{W_R}-M_{\Delta^{--}}$ histogram of the   
cross section of $e^-e^- \rightarrow
b\:b\:\bar{t}\:\bar{t} $ and the contour levels
corresponding to $\sigma =0.015\: 0.15, \:1.5$  fb
for the colliding energy $\sqrt{s} = 1$ TeV
and right-handed neutrino mass $m_{\nu_2} = 1.5$ TeV.
The histogram has the evident resonanse behavior
in $M_{\Delta^{--}}$, when $M_{W_R}$ is kept fixed,
while the increase of the charged boson mass
with $M_{\Delta^{--}}= const$ causes the decrease
of the cross section.
This happens because the
 gauge coupling
remains fixed and hence the
inrease of charged boson mass leads 
to the increase of the Higgs-Majoron VEV
which should be compensated 
(since neutrino mass is also fixed)
by the decrease of the Yukawa
coupling $h_{R,11}$.

As can been seen from these contours, near the 
$\Delta^{--}$ resonance the process will be sensitive to
values of $M_{W_R}$ that 
are much above the present lower limit of 650 GeV obtained from  direct searches at Tevatron, and that
also exceed the collision energy, assuming that
 some tens of annual events is enough for the signal. Away from
 resonance, the bound one could obtain on
 $M_{W_R}$ is about 700 GeV, i.e. no improvent to the present bound. The constraint on the 
mass of the doubly charged Higgs $\Delta^{--}$ is generally stronger than that on the 
$M_{W_R}$.

As the cross section is proportional to the mass of neutrino, the larger
$m_{\nu_2}$ the more stringent are the ensuing constraints.
 This is demonstrated in Fig. \cfig{s10n20}
where $m_{\nu_2}=2$ TeV.

Increasing the collision energy will, of course, lead to more 
restrictive bounds. In Fig. \cfig{s15n10} and 
\cfig{s15n15} we present the sensitivity contours
for $\sqrt s=1.5$ TeV with the masses of the right-handed
neutrinos 1 and 1.5 TeV, respectively. The achievable limit for
$M_{W_R}$ is now about 1.5 TeV at the triplet Higgs
resonance and outside the resonance  about 1 TeV,
 a considerable improvement to the present bound.

Let us now consider the case when the final state
 quarks are light.  
As one would expect, the results are very similar
 to the heavy quark case considered above.
     We checked this by the CompHEP calculations and found
     that 
     the relative difference in the cross section 
     is of the order of $m_t/ \sqrt s$, i.e. 20--25\%.
If we impose for the counterparts of the top 
quarks, the $c$ quarks, in the reaction $e^-e^- \rightarrow s\:s \:\bar{c}\: \bar{c}$ the  cut
$E_{\bar{c}_{1,2}} > 190$ GeV,
 which is very effective in diminishing 
 the SM background (see below), 
the cross sections differ not more than 12 \%.
  Accordingly,  we have these approximative relationships among the heavy and light quark
cases:

\be
\sigma(bb\bar{t} \bar{t})\approx
\sigma(ss\bar{c} \bar{c})\approx
\sigma(dd\bar{u} \bar{u}).
\lbl{relations}
\ee
Assuming that the Kobayashi-Maskawa
matrix elements for the right-handed currents
are the same as for the left-handed ones, the greatest non-diagonal element
is  $|V_{us}| \approx 0.221$. This will yield  a suppression factor of $4\cdot 10^{-2}$
to the cross section $\sigma(ss\bar{c} \bar{u})$ as compared with the cross sections above that contain only diagonal currents, and for the other non-diagonal processes the suppression is even stronger.
   
 In addition to the relationships \ceq{relations} one can immediately write down the following approximative relations:  
  $$
   2 \sigma(bb\bar{t} \bar{t})\approx
  \sigma(bs\bar{t} \bar{c})\approx
   \sigma(bd\bar{t} \bar{u})\approx
   \sigma(sd\bar{c}\bar{u}).
  $$
   The factor of two in front of the cross section of
    $e^-e^- \rightarrow bb \bar{t} \bar{t}$ originates in the identity of the final state quarks and antiquarks.  We have  checked also these  relations
 by CompHEP.

In conclusion, we have the following relations between the cross sections of the reactions with no, one and two $b$-jets in the final state:   
   
   \be
   \sigma(0b)\approx\sigma(1b)\approx 4 \cdot \sigma(2b); \:
   \lbl{rel}
   \ee
This relation may be very useful as a test
of the LR-model.
We show the sensitivity contours  for the processes
with
one final state b-jet and with no final state b-jets
in Fig. \cfig{s10lq} and Fig. \cfig{s15lq}, respectively.
In correspondence with \ceq{rel}
they  will yield more severe restrictions for the
$M_{W_R}$ mass than the heavy quark final state
and would give an essential improvement to the present bound.

\section{The SM background}

The main SM background of the reaction $e^-e^- \rightarrow b\:b \:\bar{t}\: \bar{t}$ is 
due to the process $e^-e^- \rightarrow  \nu_e \nu_e
b\:b \:\bar{t}\: \bar{t}$, which has the same visible particles in the final state. 
(In the case of light quarks in the final state one has the analogous background process.) Furhermore, since it is not possible to distinguish
   $b$ and $\bar{b}$  jets from each other, the  SM process 
    $e^-e^- \rightarrow  e^- e^-
   b\:t \:\bar{b}\: \bar{t}$
   (and analogously for the other quark combinations) is  another important source of 
   background. 

To analyse these processes
   we first note that  quark-antiquark
   pairs can be produced only from
 $W$, $Z$ or Higgs lines. Therefore one can start with considering the processes
 $e^-e^- \rightarrow e^-e^-W^+W^-$;   
  $e^-e^- \rightarrow e^-e^-ZZ$;
  $e^-e^- \rightarrow e^-e^-ZH$;   
    $e^-e^- \rightarrow e^-e^-HH$;
    $e^-e^- \rightarrow \nu_e\nu_eW^-W^-$,
which were analysed in
 \cite{bckgr}. The first reaction has the largest
    cross section: at $\sqrt s=1$ TeV  about
    800 fb  and at  $\sqrt s=1.5$ TeV  about
    1100 fb. All the other processes are at
    least 50 times smaller and may be neglected
    here.
    If we use for a conservative estimation
    the so-called $product \times decay$
    (or narrow width)
    approximation \cite{redbook},  which assumes
    that the intermediate $W_L^{\pm}$-bosons are mostly on shell,
    we will get for the background values
    of order 50--70 fb, which is  inconveniently high    in comparison with
our signal.
However, if we impose the cut of 50 GeV on the energies of the final state electrons (whose energy we assume to be possible to determine as missing energy allthough particles themselves may not be distinguishable
from the beam particles),
the cross section 
$\sigma(e^-e^- \rightarrow e^-\:e^-\:W^+\:W^-)$
diminishes by 3 orders of magnitude
and yields the background at 1 TeV
on the  $0.1$ fb  level
and at 1.5 TeV on the $0.03$ fb.
There is a further suppression in the case
 of  the $b b \bar{t} \bar{t}$
due to the fact the intermediate $W_L$
bosons should actually be  away from the pole
as  the invariant mass of its decay products  $b,\bar{t}$
 should be greater than $m_t$.
This yields alltogether  8 orders of magnitude
    suppression  of the background, making it fully harmless.

The situation will the the same for the light quarks
if one applies the corresponding cut on the
2-jet invariant mass, i.e. $s> m_t^2$. Thus by means
of the cut combined with the measurements of the missing
energy it is possible to make the SM background
about 7 orders of magnitude smaller than the
investigated process.

 Even without measuring the missing
energy associated with  the electrons, imposing just the cut on 
the invariant mass, it will be possible  to make the SM background 4 orders of magnitude
smaller than the investigated signal.
    
Thus we can conclude that the 
$e^-e^- \rightarrow q\:q\: \bar{q}\: \bar{q}$
processes in the LR-symmetric model 
may be well observed above the SM background.

\section{Summary}

The main results of this paper can be summarized as follows.
It is shown that the reaction $e^-e^- \rightarrow q\:q\:\bar{q}\: \bar{q}$
may be observed at NLC for a wide range of reasonable parameter values of the
left-right symmetric model and already
below the $W_R $ threshold.
For the collision energy  $\sqrt s=1.5$ TeV
and luminosity $ 10 ^{35} {\rm cm}^{-2}
\cdot {\rm s}^{-1}$ the lower limit for the mass of
the right-handed gauge boson one could reach is $M_{W_R} \gsim 1000$ GeV.
Near the doubly charged Higgs ($\Delta^{--}$) resonance the lower bound on $M_{W_R}$ 
may reach, and even exceed, the value of the collision energy.
As the lepton number violation and neutrino masses are intimately connected through the Maojaran mass terms, the strength of 
the $e^-e^- \rightarrow q\:q\:\bar{q}\: \bar{q}$
process increases with the growth of the    mass of the right-handed neutrino.
The "non-diagonal" processes, i.e. the reactions where the $\bar q q$ pair or pairs in the final state mix with fermion families, 
are essentially suppressed, while all the "diagonal"
processes have approximately the same probability.
 Process $e^-e^- 
\rightarrow b\:b\: \bar{t}\: \bar{t}$
can be identified as $b$-tagging is possible.
For the processes involving only light quarks
 or containing just one $b$-jet are
approximately  related
to this    cross section by eq. (4.6).

The SM background can be suppressed
to the level 4 orders of magnitude below the
process rate if the proper
cuts in the phase space are
applied, and it can be made even 7 orders of magnitude below the signal level    if the full energy of the event can be
reconstructed with the  accuracy of 50 GeV.

  \section{Acknowledgments}

  One of us   (N.R.)  is extremely grateful to CIMO
  organization for the financial support 
  making his stay in Helsinki possible,
 and to Theoretical Physics Division
  of the Department of Physics of Helsinki University
  for warm hospitality.
  It is also a great pleasure to thank Alexander Pukhov
  for helpful instructions for using the CompHEP package.
This work has been  supported also by the Academy of Finland under the contract
no. 40677.

\newcommand{\bi}{\bibitem}

\newpage
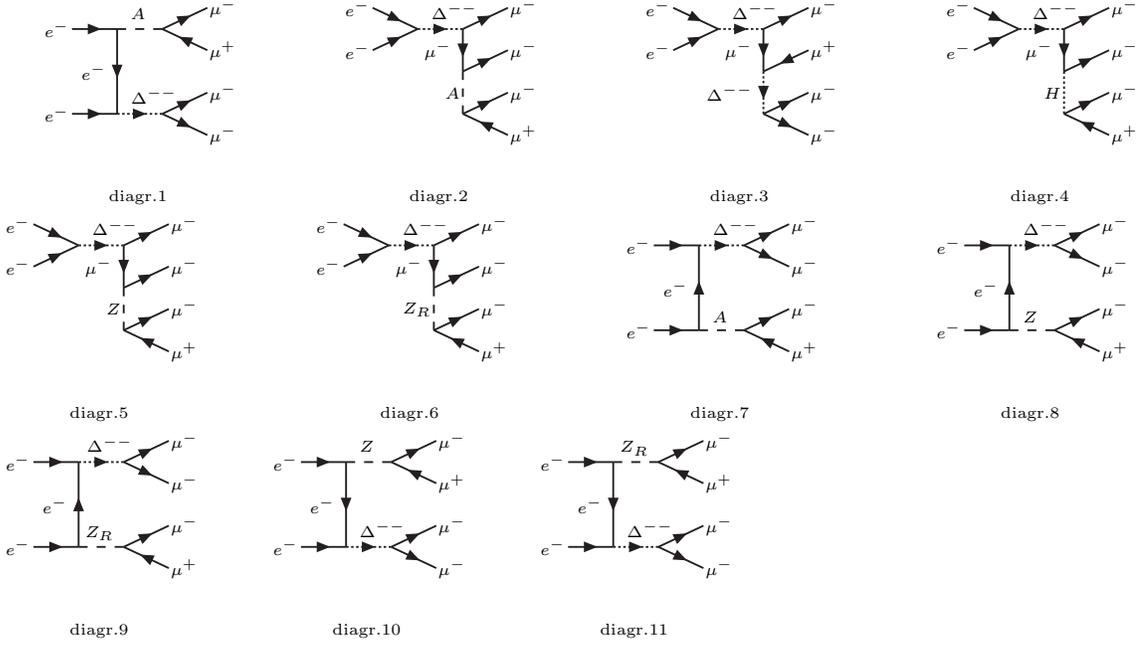
\begin{figure}[t]
{\def\chepscale{1.0} 
\unitlength=\chepscale pt
\SetWidth{0.7}      
\SetScale{\chepscale}
\tiny    
{} \qquad\allowbreak
\begin{picture}(79,81)(0,0)
\Text(13.0,65.0)[r]{$e^-$}
\ArrowLine(14.0,65.0)(31.0,65.0) 
\Text(39.0,69.0)[b]{$A$}
\DashLine(31.0,65.0)(48.0,65.0){3.0} 
\Text(66.0,73.0)[l]{$\mu^-$}
\ArrowLine(48.0,65.0)(65.0,73.0) 
\Text(66.0,57.0)[l]{$\mu^+$}
\ArrowLine(65.0,57.0)(48.0,65.0) 
\Text(27.0,49.0)[r]{$e^-$}
\ArrowLine(31.0,65.0)(31.0,33.0) 
\Text(13.0,33.0)[r]{$e^-$}
\ArrowLine(14.0,33.0)(31.0,33.0) 
\Text(45.0,37.0)[b]{$\Delta^{--}$}
\DashArrowLine(31.0,33.0)(48.0,33.0){1.0} 
\Text(66.0,41.0)[l]{$\mu^-$}
\ArrowLine(48.0,33.0)(65.0,41.0) 
\Text(66.0,25.0)[l]{$\mu^-$}
\ArrowLine(48.0,33.0)(65.0,25.0) 
\Text(39,0)[b] {diagr.1}
\end{picture} \ 
{} \qquad\allowbreak
\begin{picture}(79,81)(0,0)
\Text(13.0,73.0)[r]{$e^-$}
\ArrowLine(14.0,73.0)(31.0,65.0) 
\Text(13.0,57.0)[r]{$e^-$}
\ArrowLine(14.0,57.0)(31.0,65.0) 
\Text(45.0,69.0)[b]{$\Delta^{--}$}
\DashArrowLine(31.0,65.0)(48.0,65.0){1.0} 
\Text(66.0,73.0)[l]{$\mu^-$}
\ArrowLine(48.0,65.0)(65.0,73.0) 
\Text(44.0,57.0)[r]{$\mu^-$}
\ArrowLine(48.0,65.0)(48.0,49.0) 
\Text(66.0,57.0)[l]{$\mu^-$}
\ArrowLine(48.0,49.0)(65.0,57.0) 
\Text(47.0,41.0)[r]{$A$}
\DashLine(48.0,49.0)(48.0,33.0){3.0} 
\Text(66.0,41.0)[l]{$\mu^-$}
\ArrowLine(48.0,33.0)(65.0,41.0) 
\Text(66.0,25.0)[l]{$\mu^+$}
\ArrowLine(65.0,25.0)(48.0,33.0) 
\Text(39,0)[b] {diagr.2}
\end{picture} \ 
{} \qquad\allowbreak
\begin{picture}(79,81)(0,0)
\Text(13.0,73.0)[r]{$e^-$}
\ArrowLine(14.0,73.0)(31.0,65.0) 
\Text(13.0,57.0)[r]{$e^-$}
\ArrowLine(14.0,57.0)(31.0,65.0) 
\Text(45.0,69.0)[b]{$\Delta^{--}$}
\DashArrowLine(31.0,65.0)(48.0,65.0){1.0} 
\Text(66.0,73.0)[l]{$\mu^-$}
\ArrowLine(48.0,65.0)(65.0,73.0) 
\Text(44.0,57.0)[r]{$\mu^-$}
\ArrowLine(48.0,65.0)(48.0,49.0) 
\Text(66.0,57.0)[l]{$\mu^+$}
\ArrowLine(65.0,57.0)(48.0,49.0) 
\Text(44.0,41.0)[r]{$\Delta^{--}$}
\DashArrowLine(48.0,49.0)(48.0,33.0){1.0} 
\Text(66.0,41.0)[l]{$\mu^-$}
\ArrowLine(48.0,33.0)(65.0,41.0) 
\Text(66.0,25.0)[l]{$\mu^-$}
\ArrowLine(48.0,33.0)(65.0,25.0) 
\Text(39,0)[b] {diagr.3}
\end{picture} \ 
{} \qquad\allowbreak
\begin{picture}(79,81)(0,0)
\Text(13.0,73.0)[r]{$e^-$}
\ArrowLine(14.0,73.0)(31.0,65.0) 
\Text(13.0,57.0)[r]{$e^-$}
\ArrowLine(14.0,57.0)(31.0,65.0) 
\Text(45.0,69.0)[b]{$\Delta^{--}$}
\DashArrowLine(31.0,65.0)(48.0,65.0){1.0} 
\Text(66.0,73.0)[l]{$\mu^-$}
\ArrowLine(48.0,65.0)(65.0,73.0) 
\Text(44.0,57.0)[r]{$\mu^-$}
\ArrowLine(48.0,65.0)(48.0,49.0) 
\Text(66.0,57.0)[l]{$\mu^-$}
\ArrowLine(48.0,49.0)(65.0,57.0) 
\Text(47.0,41.0)[r]{$H$}
\DashLine(48.0,49.0)(48.0,33.0){1.0}
\Text(66.0,41.0)[l]{$\mu^-$}
\ArrowLine(48.0,33.0)(65.0,41.0) 
\Text(66.0,25.0)[l]{$\mu^+$}
\ArrowLine(65.0,25.0)(48.0,33.0) 
\Text(39,0)[b] {diagr.4}
\end{picture} \ 
{} \qquad\allowbreak
\begin{picture}(79,81)(0,0)
\Text(13.0,73.0)[r]{$e^-$}
\ArrowLine(14.0,73.0)(31.0,65.0) 
\Text(13.0,57.0)[r]{$e^-$}
\ArrowLine(14.0,57.0)(31.0,65.0) 
\Text(45.0,69.0)[b]{$\Delta^{--}$}
\DashArrowLine(31.0,65.0)(48.0,65.0){1.0} 
\Text(66.0,73.0)[l]{$\mu^-$}
\ArrowLine(48.0,65.0)(65.0,73.0) 
\Text(44.0,57.0)[r]{$\mu^-$}
\ArrowLine(48.0,65.0)(48.0,49.0) 
\Text(66.0,57.0)[l]{$\mu^-$}
\ArrowLine(48.0,49.0)(65.0,57.0) 
\Text(47.0,41.0)[r]{$Z$}
\DashLine(48.0,49.0)(48.0,33.0){3.0} 
\Text(66.0,41.0)[l]{$\mu^-$}
\ArrowLine(48.0,33.0)(65.0,41.0) 
\Text(66.0,25.0)[l]{$\mu^+$}
\ArrowLine(65.0,25.0)(48.0,33.0) 
\Text(39,0)[b] {diagr.5}
\end{picture} \ 
{} \qquad\allowbreak
\begin{picture}(79,81)(0,0)
\Text(13.0,73.0)[r]{$e^-$}
\ArrowLine(14.0,73.0)(31.0,65.0) 
\Text(13.0,57.0)[r]{$e^-$}
\ArrowLine(14.0,57.0)(31.0,65.0) 
\Text(45.0,69.0)[b]{$\Delta^{--}$}
\DashArrowLine(31.0,65.0)(48.0,65.0){1.0} 
\Text(66.0,73.0)[l]{$\mu^-$}
\ArrowLine(48.0,65.0)(65.0,73.0) 
\Text(44.0,57.0)[r]{$\mu^-$}
\ArrowLine(48.0,65.0)(48.0,49.0) 
\Text(66.0,57.0)[l]{$\mu^-$}
\ArrowLine(48.0,49.0)(65.0,57.0) 
\Text(47.0,41.0)[r]{$Z_R$}
\DashLine(48.0,49.0)(48.0,33.0){3.0} 
\Text(66.0,41.0)[l]{$\mu^-$}
\ArrowLine(48.0,33.0)(65.0,41.0) 
\Text(66.0,25.0)[l]{$\mu^+$}
\ArrowLine(65.0,25.0)(48.0,33.0) 
\Text(39,0)[b] {diagr.6}
\end{picture} \ 
{} \qquad\allowbreak
\begin{picture}(79,81)(0,0)
\Text(13.0,65.0)[r]{$e^-$}
\ArrowLine(14.0,65.0)(31.0,65.0) 
\Text(45.0,69.0)[b]{$\Delta^{--}$}
\DashArrowLine(31.0,65.0)(48.0,65.0){1.0} 
\Text(66.0,73.0)[l]{$\mu^-$}
\ArrowLine(48.0,65.0)(65.0,73.0) 
\Text(66.0,57.0)[l]{$\mu^-$}
\ArrowLine(48.0,65.0)(65.0,57.0) 
\Text(27.0,49.0)[r]{$e^-$}
\ArrowLine(31.0,33.0)(31.0,65.0) 
\Text(13.0,33.0)[r]{$e^-$}
\ArrowLine(14.0,33.0)(31.0,33.0) 
\Text(39.0,36.0)[b]{$A$}
\DashLine(31.0,33.0)(48.0,33.0){3.0} 
\Text(66.0,41.0)[l]{$\mu^-$}
\ArrowLine(48.0,33.0)(65.0,41.0) 
\Text(66.0,25.0)[l]{$\mu^+$}
\ArrowLine(65.0,25.0)(48.0,33.0) 
\Text(39,0)[b] {diagr.7}
\end{picture} \ 
{} \qquad\allowbreak
\begin{picture}(79,81)(0,0)
\Text(13.0,65.0)[r]{$e^-$}
\ArrowLine(14.0,65.0)(31.0,65.0) 
\Text(45.0,69.0)[b]{$\Delta^{--}$}
\DashArrowLine(31.0,65.0)(48.0,65.0){1.0} 
\Text(66.0,73.0)[l]{$\mu^-$}
\ArrowLine(48.0,65.0)(65.0,73.0) 
\Text(66.0,57.0)[l]{$\mu^-$}
\ArrowLine(48.0,65.0)(65.0,57.0) 
\Text(27.0,49.0)[r]{$e^-$}
\ArrowLine(31.0,33.0)(31.0,65.0) 
\Text(13.0,33.0)[r]{$e^-$}
\ArrowLine(14.0,33.0)(31.0,33.0) 
\Text(39.0,36.0)[b]{$Z$}
\DashLine(31.0,33.0)(48.0,33.0){3.0} 
\Text(66.0,41.0)[l]{$\mu^-$}
\ArrowLine(48.0,33.0)(65.0,41.0) 
\Text(66.0,25.0)[l]{$\mu^+$}
\ArrowLine(65.0,25.0)(48.0,33.0) 
\Text(39,0)[b] {diagr.8}
\end{picture} \ 
{} \qquad\allowbreak
\begin{picture}(79,81)(0,0)
\Text(13.0,65.0)[r]{$e^-$}
\ArrowLine(14.0,65.0)(31.0,65.0) 
\Text(43.0,69.0)[b]{$\Delta^{--}$}
\DashArrowLine(31.0,65.0)(48.0,65.0){1.0} 
\Text(66.0,73.0)[l]{$\mu^-$}
\ArrowLine(48.0,65.0)(65.0,73.0) 
\Text(66.0,57.0)[l]{$\mu^-$}
\ArrowLine(48.0,65.0)(65.0,57.0) 
\Text(27.0,49.0)[r]{$e^-$}
\ArrowLine(31.0,33.0)(31.0,65.0) 
\Text(13.0,33.0)[r]{$e^-$}
\ArrowLine(14.0,33.0)(31.0,33.0) 
\Text(39.0,37.0)[b]{$Z_R$}
\DashLine(31.0,33.0)(48.0,33.0){3.0} 
\Text(66.0,41.0)[l]{$\mu^-$}
\ArrowLine(48.0,33.0)(65.0,41.0) 
\Text(66.0,25.0)[l]{$\mu^+$}
\ArrowLine(65.0,25.0)(48.0,33.0) 
\Text(39,0)[b] {diagr.9}
\end{picture} \ 
{} \qquad\allowbreak
\begin{picture}(79,81)(0,0)
\Text(13.0,65.0)[r]{$e^-$}
\ArrowLine(14.0,65.0)(31.0,65.0) 
\Text(39.0,69.0)[b]{$Z$}
\DashLine(31.0,65.0)(48.0,65.0){3.0} 
\Text(66.0,73.0)[l]{$\mu^-$}
\ArrowLine(48.0,65.0)(65.0,73.0) 
\Text(66.0,57.0)[l]{$\mu^+$}
\ArrowLine(65.0,57.0)(48.0,65.0) 
\Text(27.0,49.0)[r]{$e^-$}
\ArrowLine(31.0,65.0)(31.0,33.0) 
\Text(13.0,33.0)[r]{$e^-$}
\ArrowLine(14.0,33.0)(31.0,33.0) 
\Text(45.0,37.0)[b]{$\Delta^{--}$}
\DashArrowLine(31.0,33.0)(48.0,33.0){1.0} 
\Text(66.0,41.0)[l]{$\mu^-$}
\ArrowLine(48.0,33.0)(65.0,41.0) 
\Text(66.0,25.0)[l]{$\mu^-$}
\ArrowLine(48.0,33.0)(65.0,25.0) 
\Text(39,0)[b] {diagr.10}
\end{picture} \ 
{} \qquad\allowbreak
\begin{picture}(79,81)(0,0)
\Text(13.0,65.0)[r]{$e^-$}
\ArrowLine(14.0,65.0)(31.0,65.0) 
\Text(39.0,69.0)[b]{$Z_R$}
\DashLine(31.0,65.0)(48.0,65.0){3.0} 
\Text(66.0,73.0)[l]{$\mu^-$}
\ArrowLine(48.0,65.0)(65.0,73.0) 
\Text(66.0,57.0)[l]{$\mu^+$}
\ArrowLine(65.0,57.0)(48.0,65.0) 
\Text(27.0,49.0)[r]{$e^-$}
\ArrowLine(31.0,65.0)(31.0,33.0) 
\Text(13.0,33.0)[r]{$e^-$}
\ArrowLine(14.0,33.0)(31.0,33.0) 
\Text(45.0,37.0)[b]{$\Delta^{--}$}
\DashArrowLine(31.0,33.0)(48.0,33.0){1.0} 
\Text(66.0,41.0)[l]{$\mu^-$}
\ArrowLine(48.0,33.0)(65.0,41.0) 
\Text(66.0,25.0)[l]{$\mu^-$}
\ArrowLine(48.0,33.0)(65.0,25.0) 
\Text(39,0)[b] {diagr.11}
\end{picture} \ 
}
\caption{Feynman diagramms for $e^-e^- \rightarrow
\mu^-\mu^-\mu^-\mu^+$ in the LR-model.}
\end{figure}
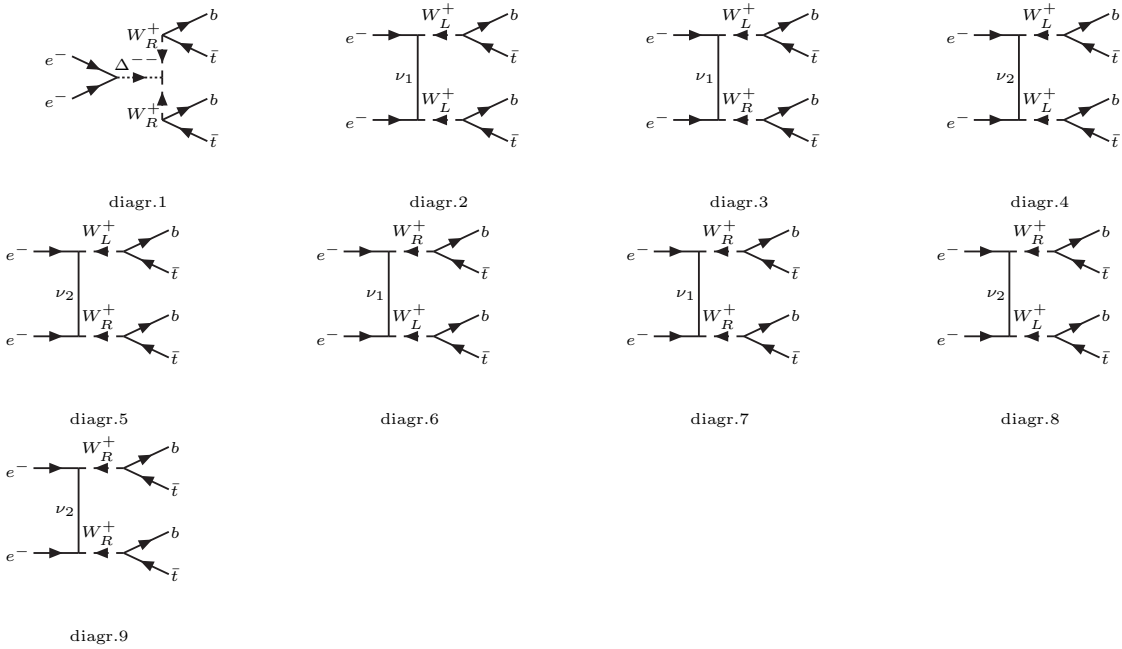
\begin{figure}[t]
{\def\chepscale{1.0} 
\unitlength=\chepscale pt
\SetWidth{0.7}      
\SetScale{\chepscale}
\tiny    
{} \qquad\allowbreak
\begin{picture}(79,81)(0,0)
\Text(13.0,57.0)[r]{$e^-$}
\ArrowLine(14.0,57.0)(31.0,49.0) 
\Text(13.0,41.0)[r]{$e^-$}
\ArrowLine(14.0,41.0)(31.0,49.0) 
\Text(30.0,52.0)[lb]{$\Delta^{--}$}
\DashArrowLine(31.0,49.0)(48.0,49.0){1.0} 
\Text(48.0,65.0)[r]{$W_R^+$}
\DashArrowLine(48.0,65.0)(48.0,49.0){3.0} 
\Text(66.0,73.0)[l]{$b$}
\ArrowLine(48.0,65.0)(65.0,73.0) 
\Text(66.0,57.0)[l]{$\bar{t}$}
\ArrowLine(65.0,57.0)(48.0,65.0) 
\Text(48.0,35.0)[r]{$W_R^+$}
\DashArrowLine(48.0,33.0)(48.0,49.0){3.0} 
\Text(66.0,41.0)[l]{$b$}
\ArrowLine(48.0,33.0)(65.0,41.0) 
\Text(66.0,25.0)[l]{$\bar{t}$}
\ArrowLine(65.0,25.0)(48.0,33.0) 
\Text(39,0)[b] {diagr.1}
\end{picture} \ 
{} \qquad\allowbreak
\begin{picture}(79,81)(0,0)
\Text(13.0,65.0)[r]{$e^-$}
\ArrowLine(14.0,65.0)(31.0,65.0) 
\Text(39.0,71.0)[b]{$W_L^+$}
\DashArrowLine(48.0,65.0)(31.0,65.0){3.0} 
\Text(66.0,73.0)[l]{$b$}
\ArrowLine(48.0,65.0)(65.0,73.0) 
\Text(66.0,57.0)[l]{$\bar{t}$}
\ArrowLine(65.0,57.0)(48.0,65.0) 
\Text(30.0,49.0)[r]{$\nu_1$}
\Line(31.0,65.0)(31.0,33.0) 
\Text(13.0,33.0)[r]{$e^-$}
\ArrowLine(14.0,33.0)(31.0,33.0) 
\Text(39.0,39.0)[b]{$W_L^+$}
\DashArrowLine(48.0,33.0)(31.0,33.0){3.0} 
\Text(66.0,41.0)[l]{$b$}
\ArrowLine(48.0,33.0)(65.0,41.0) 
\Text(66.0,25.0)[l]{$\bar{t}$}
\ArrowLine(65.0,25.0)(48.0,33.0) 
\Text(39,0)[b] {diagr.2}
\end{picture} \ 
{} \qquad\allowbreak
\begin{picture}(79,81)(0,0)
\Text(13.0,65.0)[r]{$e^-$}
\ArrowLine(14.0,65.0)(31.0,65.0) 
\Text(39.0,71.0)[b]{$W_L^+$}
\DashArrowLine(48.0,65.0)(31.0,65.0){3.0} 
\Text(66.0,73.0)[l]{$b$}
\ArrowLine(48.0,65.0)(65.0,73.0) 
\Text(66.0,57.0)[l]{$\bar{t}$}
\ArrowLine(65.0,57.0)(48.0,65.0) 
\Text(30.0,49.0)[r]{$\nu_1$}
\Line(31.0,65.0)(31.0,33.0) 
\Text(13.0,33.0)[r]{$e^-$}
\ArrowLine(14.0,33.0)(31.0,33.0) 
\Text(39.0,39.0)[b]{$W_R^+$}
\DashArrowLine(48.0,33.0)(31.0,33.0){3.0} 
\Text(66.0,41.0)[l]{$b$}
\ArrowLine(48.0,33.0)(65.0,41.0) 
\Text(66.0,25.0)[l]{$\bar{t}$}
\ArrowLine(65.0,25.0)(48.0,33.0) 
\Text(39,0)[b] {diagr.3}
\end{picture} \ 
{} \qquad\allowbreak
\begin{picture}(79,81)(0,0)
\Text(13.0,65.0)[r]{$e^-$}
\ArrowLine(14.0,65.0)(31.0,65.0) 
\Text(39.0,71.0)[b]{$W_L^+$}
\DashArrowLine(48.0,65.0)(31.0,65.0){3.0} 
\Text(66.0,73.0)[l]{$b$}
\ArrowLine(48.0,65.0)(65.0,73.0) 
\Text(66.0,57.0)[l]{$\bar{t}$}
\ArrowLine(65.0,57.0)(48.0,65.0) 
\Text(30.0,49.0)[r]{$\nu_2$}
\Line(31.0,65.0)(31.0,33.0) 
\Text(13.0,33.0)[r]{$e^-$}
\ArrowLine(14.0,33.0)(31.0,33.0) 
\Text(39.0,39.0)[b]{$W_L^+$}
\DashArrowLine(48.0,33.0)(31.0,33.0){3.0} 
\Text(66.0,41.0)[l]{$b$}
\ArrowLine(48.0,33.0)(65.0,41.0) 
\Text(66.0,25.0)[l]{$\bar{t}$}
\ArrowLine(65.0,25.0)(48.0,33.0) 
\Text(39,0)[b] {diagr.4}
\end{picture} \ 
{} \qquad\allowbreak
\begin{picture}(79,81)(0,0)
\Text(13.0,65.0)[r]{$e^-$}
\ArrowLine(14.0,65.0)(31.0,65.0) 
\Text(39.0,71.0)[b]{$W_L^+$}
\DashArrowLine(48.0,65.0)(31.0,65.0){3.0} 
\Text(66.0,73.0)[l]{$b$}
\ArrowLine(48.0,65.0)(65.0,73.0) 
\Text(66.0,57.0)[l]{$\bar{t}$}
\ArrowLine(65.0,57.0)(48.0,65.0) 
\Text(30.0,49.0)[r]{$\nu_2$}
\Line(31.0,65.0)(31.0,33.0) 
\Text(13.0,33.0)[r]{$e^-$}
\ArrowLine(14.0,33.0)(31.0,33.0) 
\Text(39.0,39.0)[b]{$W_R^+$}
\DashArrowLine(48.0,33.0)(31.0,33.0){3.0} 
\Text(66.0,41.0)[l]{$b$}
\ArrowLine(48.0,33.0)(65.0,41.0) 
\Text(66.0,25.0)[l]{$\bar{t}$}
\ArrowLine(65.0,25.0)(48.0,33.0) 
\Text(39,0)[b] {diagr.5}
\end{picture} \ 
{} \qquad\allowbreak
\begin{picture}(79,81)(0,0)
\Text(13.0,65.0)[r]{$e^-$}
\ArrowLine(14.0,65.0)(31.0,65.0) 
\Text(39.0,71.0)[b]{$W_R^+$}
\DashArrowLine(48.0,65.0)(31.0,65.0){3.0} 
\Text(66.0,73.0)[l]{$b$}
\ArrowLine(48.0,65.0)(65.0,73.0) 
\Text(66.0,57.0)[l]{$\bar{t}$}
\ArrowLine(65.0,57.0)(48.0,65.0) 
\Text(30.0,49.0)[r]{$\nu_1$}
\Line(31.0,65.0)(31.0,33.0) 
\Text(13.0,33.0)[r]{$e^-$}
\ArrowLine(14.0,33.0)(31.0,33.0) 
\Text(39.0,39.0)[b]{$W_L^+$}
\DashArrowLine(48.0,33.0)(31.0,33.0){3.0} 
\Text(66.0,41.0)[l]{$b$}
\ArrowLine(48.0,33.0)(65.0,41.0) 
\Text(66.0,25.0)[l]{$\bar{t}$}
\ArrowLine(65.0,25.0)(48.0,33.0) 
\Text(39,0)[b] {diagr.6}
\end{picture} \ 
{} \qquad\allowbreak
\begin{picture}(79,81)(0,0)
\Text(13.0,65.0)[r]{$e^-$}
\ArrowLine(14.0,65.0)(31.0,65.0) 
\Text(39.0,71.0)[b]{$W_R^+$}
\DashArrowLine(48.0,65.0)(31.0,65.0){3.0} 
\Text(66.0,73.0)[l]{$b$}
\ArrowLine(48.0,65.0)(65.0,73.0) 
\Text(66.0,57.0)[l]{$\bar{t}$}
\ArrowLine(65.0,57.0)(48.0,65.0) 
\Text(30.0,49.0)[r]{$\nu_1$}
\Line(31.0,65.0)(31.0,33.0) 
\Text(13.0,33.0)[r]{$e^-$}
\ArrowLine(14.0,33.0)(31.0,33.0) 
\Text(39.0,39.0)[b]{$W_R^+$}
\DashArrowLine(48.0,33.0)(31.0,33.0){3.0} 
\Text(66.0,41.0)[l]{$b$}
\ArrowLine(48.0,33.0)(65.0,41.0) 
\Text(66.0,25.0)[l]{$\bar{t}$}
\ArrowLine(65.0,25.0)(48.0,33.0) 
\Text(39,0)[b] {diagr.7}
\end{picture} \ 
{} \qquad\allowbreak
\begin{picture}(79,81)(0,0)
\Text(13.0,65.0)[r]{$e^-$}
\ArrowLine(14.0,65.0)(31.0,65.0) 
\Text(39.0,71.0)[b]{$W_R^+$}
\DashArrowLine(48.0,65.0)(31.0,65.0){3.0} 
\Text(66.0,73.0)[l]{$b$}
\ArrowLine(48.0,65.0)(65.0,73.0) 
\Text(66.0,57.0)[l]{$\bar{t}$}
\ArrowLine(65.0,57.0)(48.0,65.0) 
\Text(30.0,49.0)[r]{$\nu_2$}
\Line(31.0,65.0)(31.0,33.0) 
\Text(13.0,33.0)[r]{$e^-$}
\ArrowLine(14.0,33.0)(31.0,33.0) 
\Text(39.0,39.0)[b]{$W_L^+$}
\DashArrowLine(48.0,33.0)(31.0,33.0){3.0} 
\Text(66.0,41.0)[l]{$b$}
\ArrowLine(48.0,33.0)(65.0,41.0) 
\Text(66.0,25.0)[l]{$\bar{t}$}
\ArrowLine(65.0,25.0)(48.0,33.0) 
\Text(39,0)[b] {diagr.8}
\end{picture} \ 
{} \qquad\allowbreak
\begin{picture}(79,81)(0,0)
\Text(13.0,65.0)[r]{$e^-$}
\ArrowLine(14.0,65.0)(31.0,65.0) 
\Text(39.0,71.0)[b]{$W_R^+$}
\DashArrowLine(48.0,65.0)(31.0,65.0){3.0} 
\Text(66.0,73.0)[l]{$b$}
\ArrowLine(48.0,65.0)(65.0,73.0) 
\Text(66.0,57.0)[l]{$\bar{t}$}
\ArrowLine(65.0,57.0)(48.0,65.0) 
\Text(30.0,49.0)[r]{$\nu_2$}
\Line(31.0,65.0)(31.0,33.0) 
\Text(13.0,33.0)[r]{$e^-$}
\ArrowLine(14.0,33.0)(31.0,33.0) 
\Text(39.0,39.0)[b]{$W_R^+$}
\DashArrowLine(48.0,33.0)(31.0,33.0){3.0} 
\Text(66.0,41.0)[l]{$b$}
\ArrowLine(48.0,33.0)(65.0,41.0) 
\Text(66.0,25.0)[l]{$\bar{t}$}
\ArrowLine(65.0,25.0)(48.0,33.0) 
\Text(39,0)[b] {diagr.9}
\end{picture} \ 
}
\caption{Feynman diagramms for $e^-e^- \rightarrow
b\:b\:\bar{t}\:\bar{t} $ in the LR-model.}
\end{figure}

\newpage

\thispagestyle{empty}

\begin{figure}[t]
\vspace*{-4.3cm}
\hspace*{-1.5cm}
\epsfysize=22cm
\epsffile{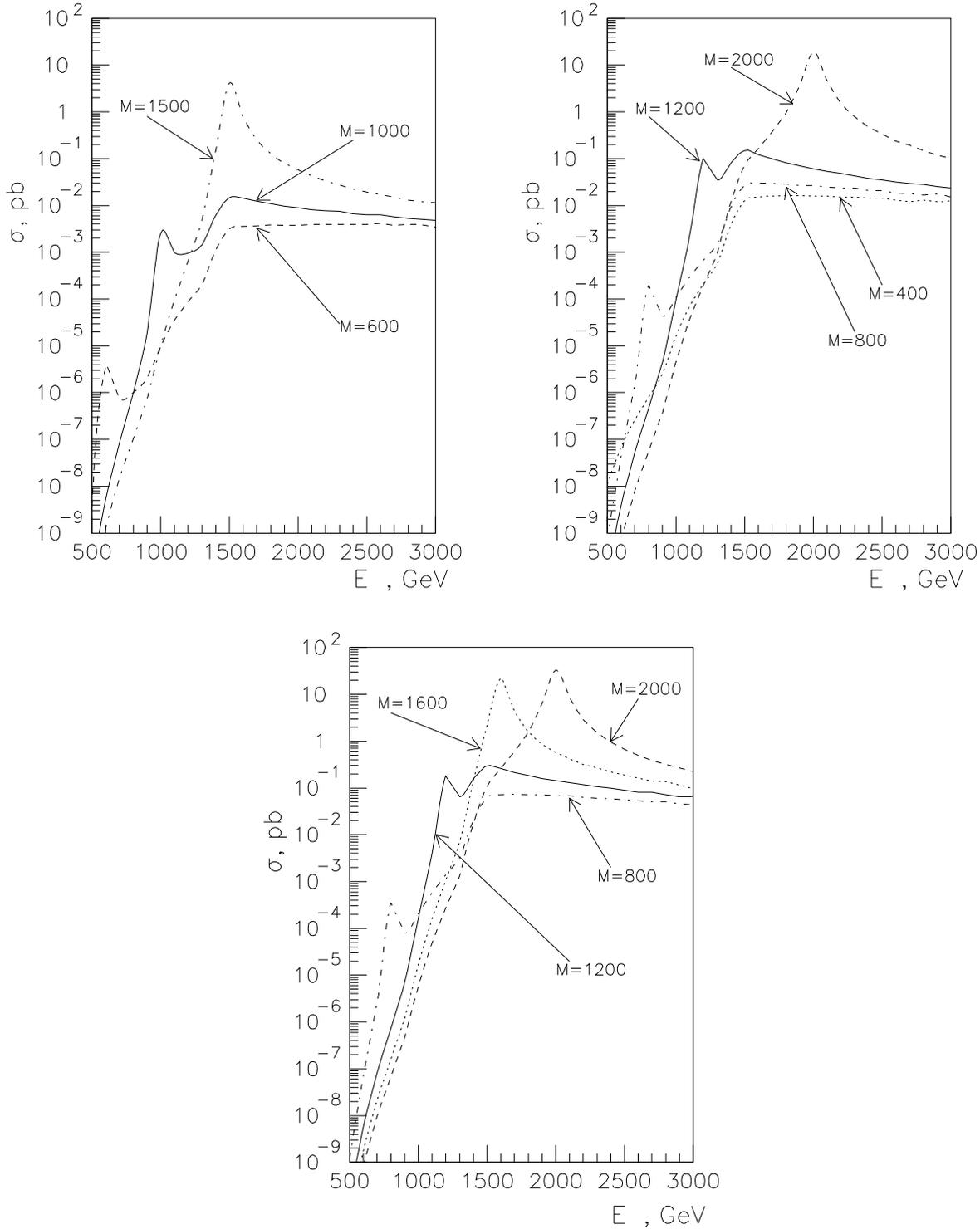}
\caption{ Energy dependence
of the full cross section for the process
$e^-e^- \rightarrow b\:b\: \bar{t}\: \bar{t}$ 
  for  different values
 of $\Delta^{--}$ mass
 ($M \equiv M_{\Delta^{--}}$)
 and right-handed neutrino masses:
 $m_{\nu_2}=1$ TeV (left upper picture),
 $m_{\nu_2}=1.5$ TeV (right upper picture),
 $m_{\nu_2}=2$ TeV (lower picture),
 (see comments in the text).}
\label{fe}
\end{figure}

\begin{figure}[t]
\vspace*{-4.3cm}
\hspace*{-0.2cm}
\epsfysize=22cm
\epsffile{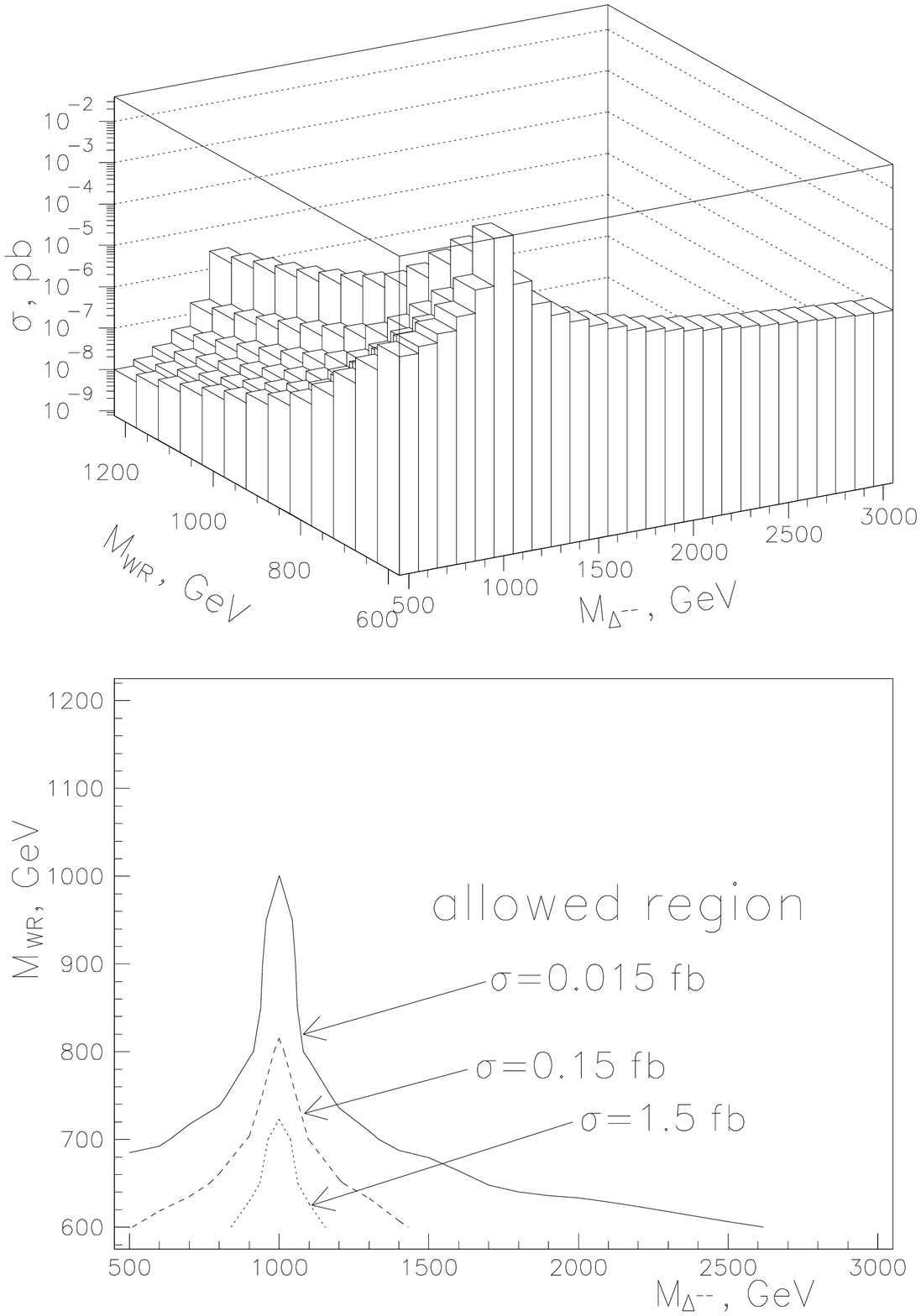}
\caption{Cross section for the process
$e^-e^- \rightarrow b\:b\: \bar{t}\: \bar{t}$
and and contours corresponding to the sensitivity levels  
$\sigma=0.01 \: fb$
\mbox{(30 events per year),}
$\sigma=0.1 \: fb$
\mbox{(300 events per year),}
$\sigma=1 \: fb$
\mbox{(3000 events per year),} 
 for the energy $E=1 \: $TeV,
 and right-handed neutrino mass
   $m_{\nu_2}=1.5 \: $TeV.}
\label{s10n15}
\end{figure}

\newpage

\begin{figure}[t]
\vspace*{-4.3cm}
\hspace*{-0.2cm}
\epsfysize=22cm
\epsffile{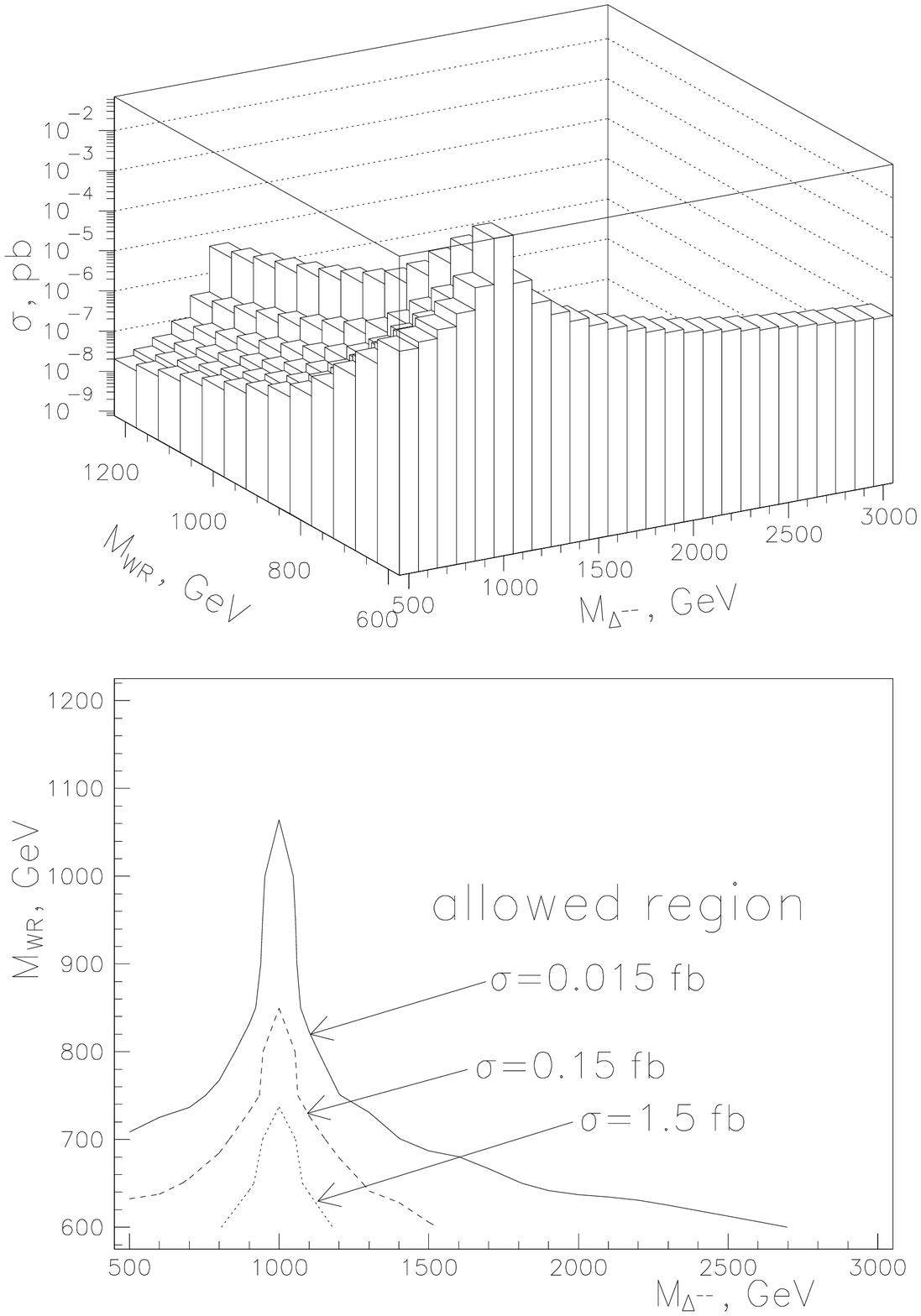}
\caption{Cross section for the process
$e^-e^- \rightarrow b\:b\:\bar{t}\: \bar{t}$
and contours corresponding to the sensitivity levels  
$\sigma=0.01 \: $ fb
\mbox{(30 events per year),}
$\sigma=0.1 \: $ fb
\mbox{(300 events per year),}
$\sigma=1 \: $ fb
\mbox{(3000 events per year),} 
 for the energy $E=1 \:$ TeV,
 and right-handed neutrino mass
   $m_{\nu_2 }=2 \:$ TeV.}
\label{s10n20}
\end{figure}

\newpage 
 
\thispagestyle{empty}

\begin{figure}[t]
\vspace*{-4.3cm}
\hspace*{-0.2cm}
\epsfysize=22cm
\epsffile{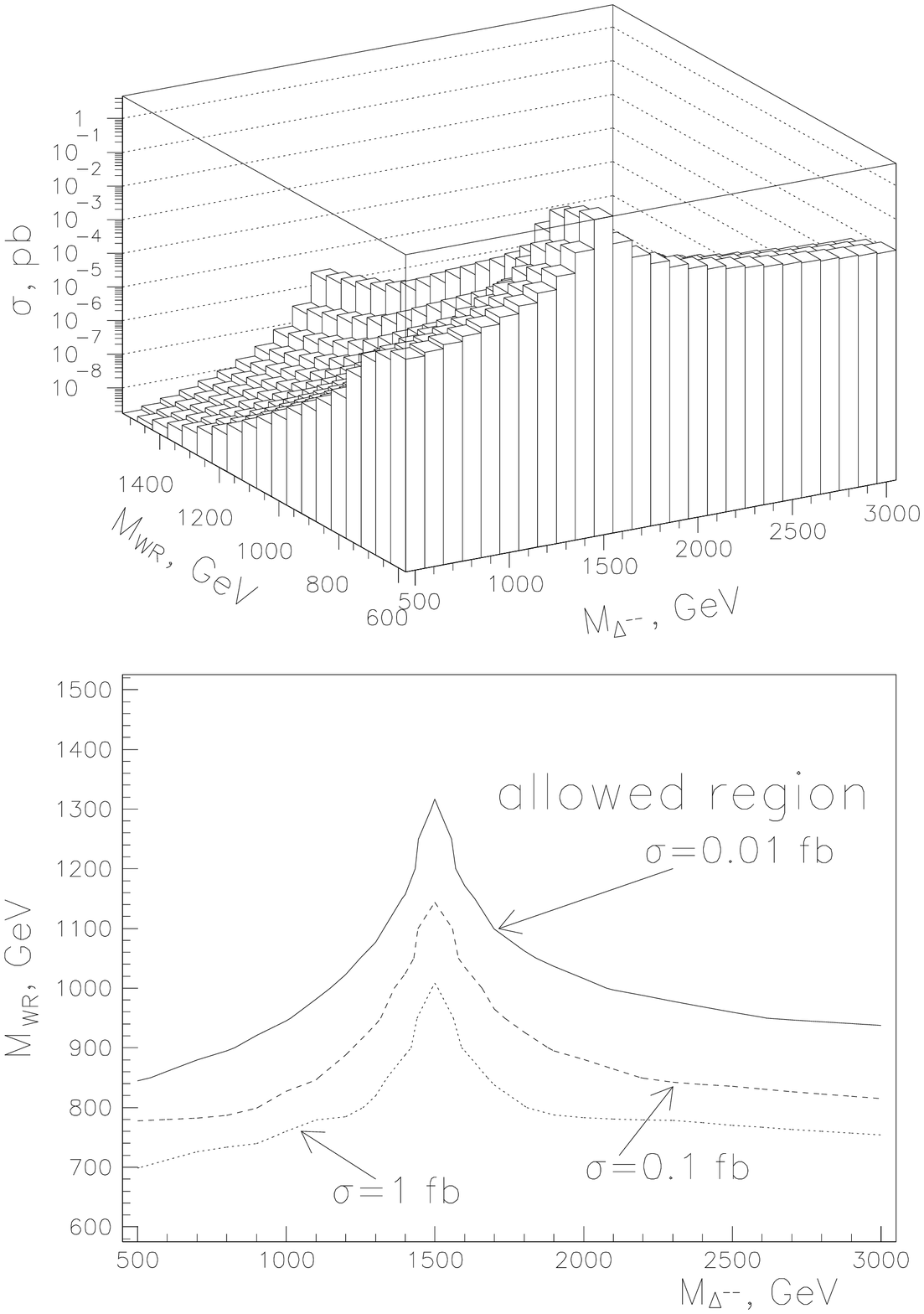}
\caption{Cross section 
for the $e^-e^- \rightarrow b,b,\bar{t}, \bar{t}$
and it's contourlevels
at $\sigma=0.015 \: $ fb
(30 events per year),
 $\sigma=0.15 \: $ fb
(300 events per year), 
at $\sigma=1.5 \: $ fb
(3000 events per year)  
for the energy $E=1.5 \: TeV $, 
and the right-handed neutrino mass
$m_{\nu_2}=1 \: $ TeV. }
\label{s15n10}
\end{figure}

\newpage

\thispagestyle{empty}

\begin{figure}[t]
\vspace*{-4.3cm}
\hspace*{-0.2cm}
\epsfysize=22cm
\epsffile{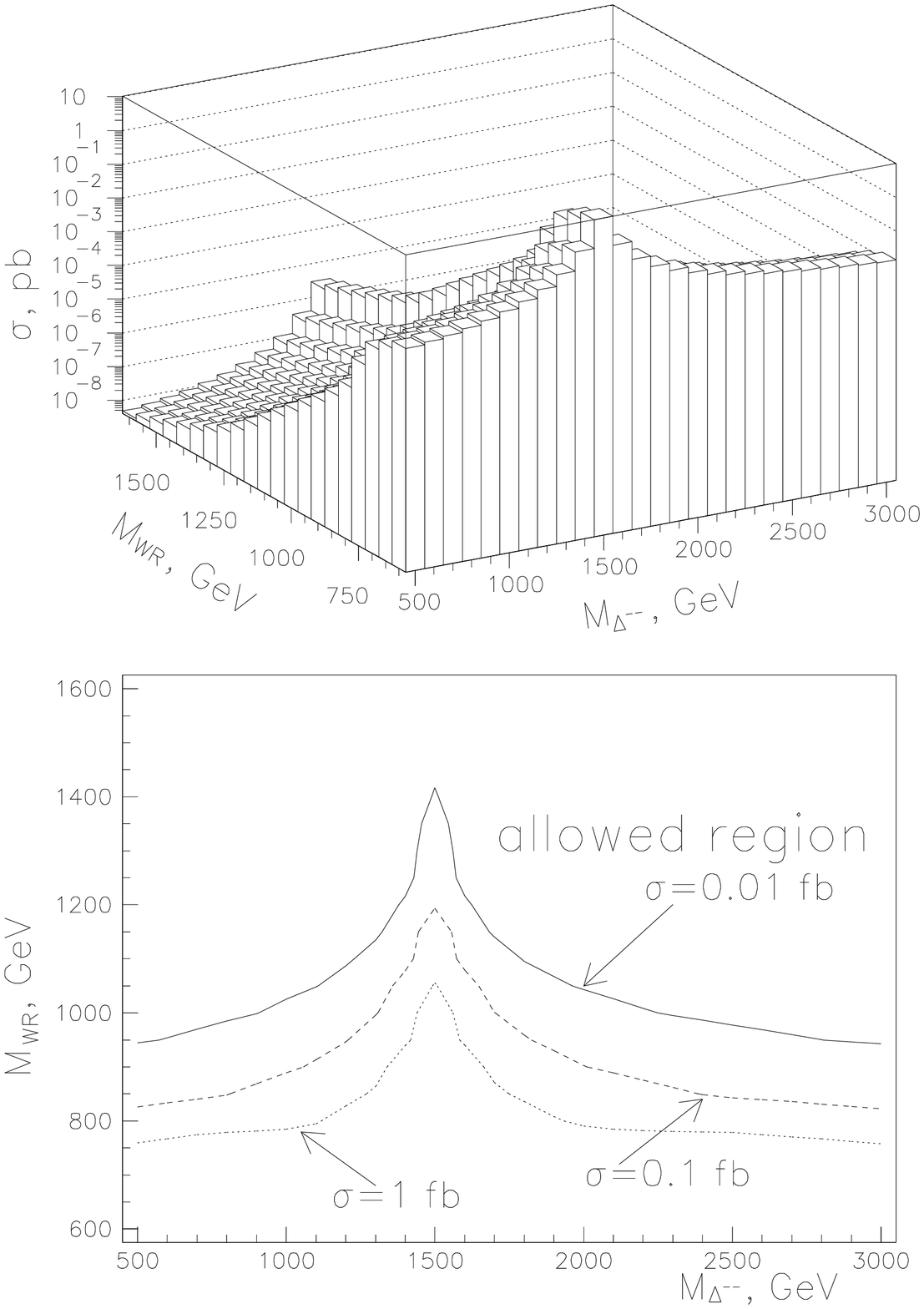}
\caption{Cross section 
for the $e^-e^- \rightarrow b,b,\bar{t}, \bar{t}$
and it's contourlevels
at $\sigma=0.01 \: $ fb
(30 events per year),
 $\sigma=0.1 \: $ fb
(300 events per year), 
 $\sigma=1 \: $ fb
(3000 events per year)  
for the energy $E=1.5 \: TeV $, 
and the right-handed neutrino mass
$m_{\nu_2}=1.5 \: $ TeV. }
\label{s15n15}
\end{figure}

 

  
\begin{figure}[t]
\vspace*{-4.3cm}
\hspace*{-0.2cm}
\epsfysize=22cm
\epsffile{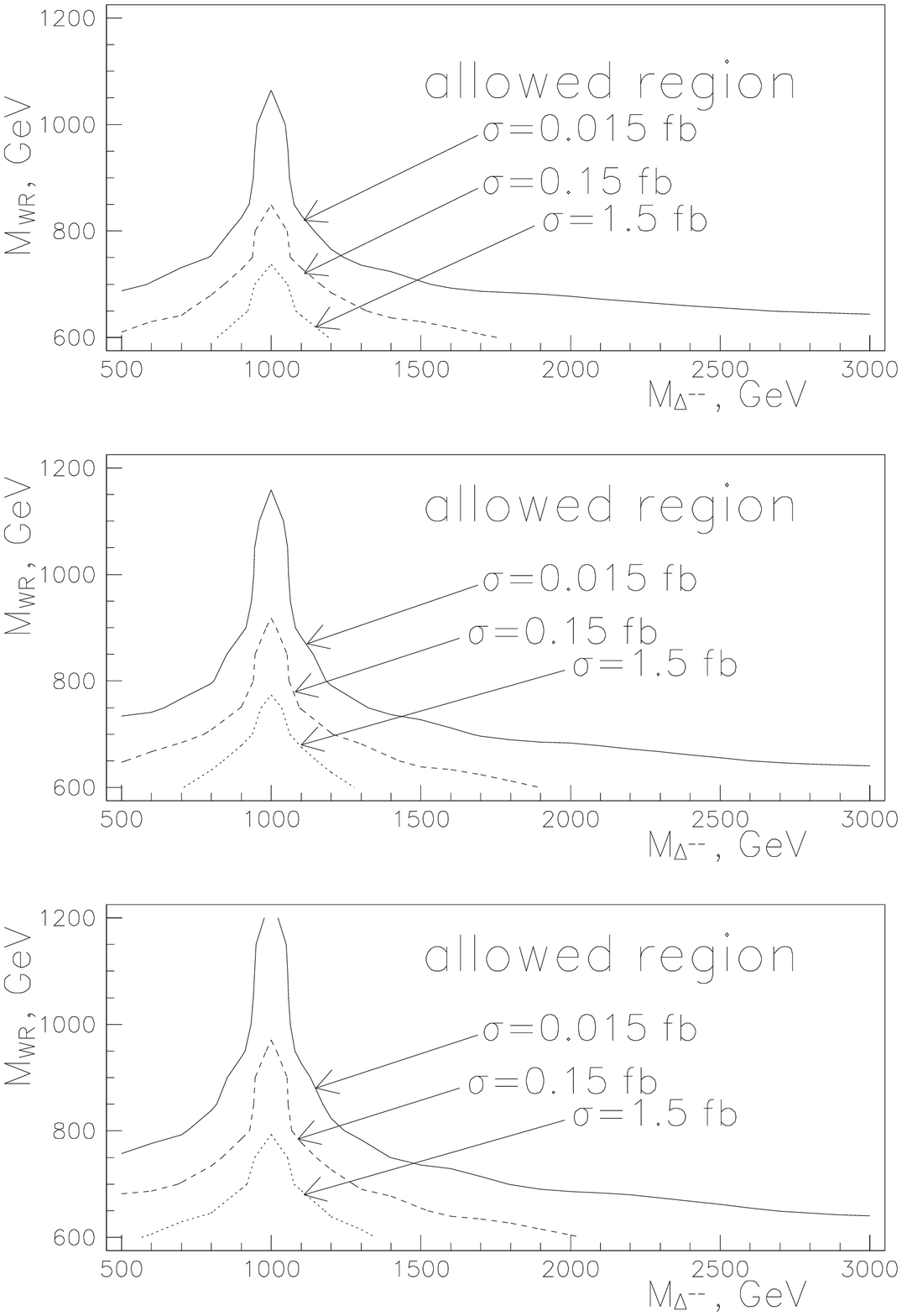}
\caption{
Contourlevels
at $\sigma=0.015 \; $ fb
(30 events per year),
 $\sigma=0.15 \; $ fb
(300 events per year), 
 $\sigma=1.5 \; $ fb
(3000 events per year)
for the processes with 1 $b$-jet
or with light-quarks only in the final state
(see comments in the text)  
for the energy $E=1 \; TeV $, 
and the right-handed neutrino masses:
$m_{\nu_2}=1 \; $ TeV  (on the top),
$m_{\nu_2}=1.5 \; $ TeV  (in the middle)
and
$m_{\nu_2}=2 \; $ TeV  (in the bottom).}
\label{s10lq}
\end{figure}

\begin{figure}[t]
\vspace*{-4.3cm}
\hspace*{-0.2cm}
\epsfysize=22cm
\epsffile{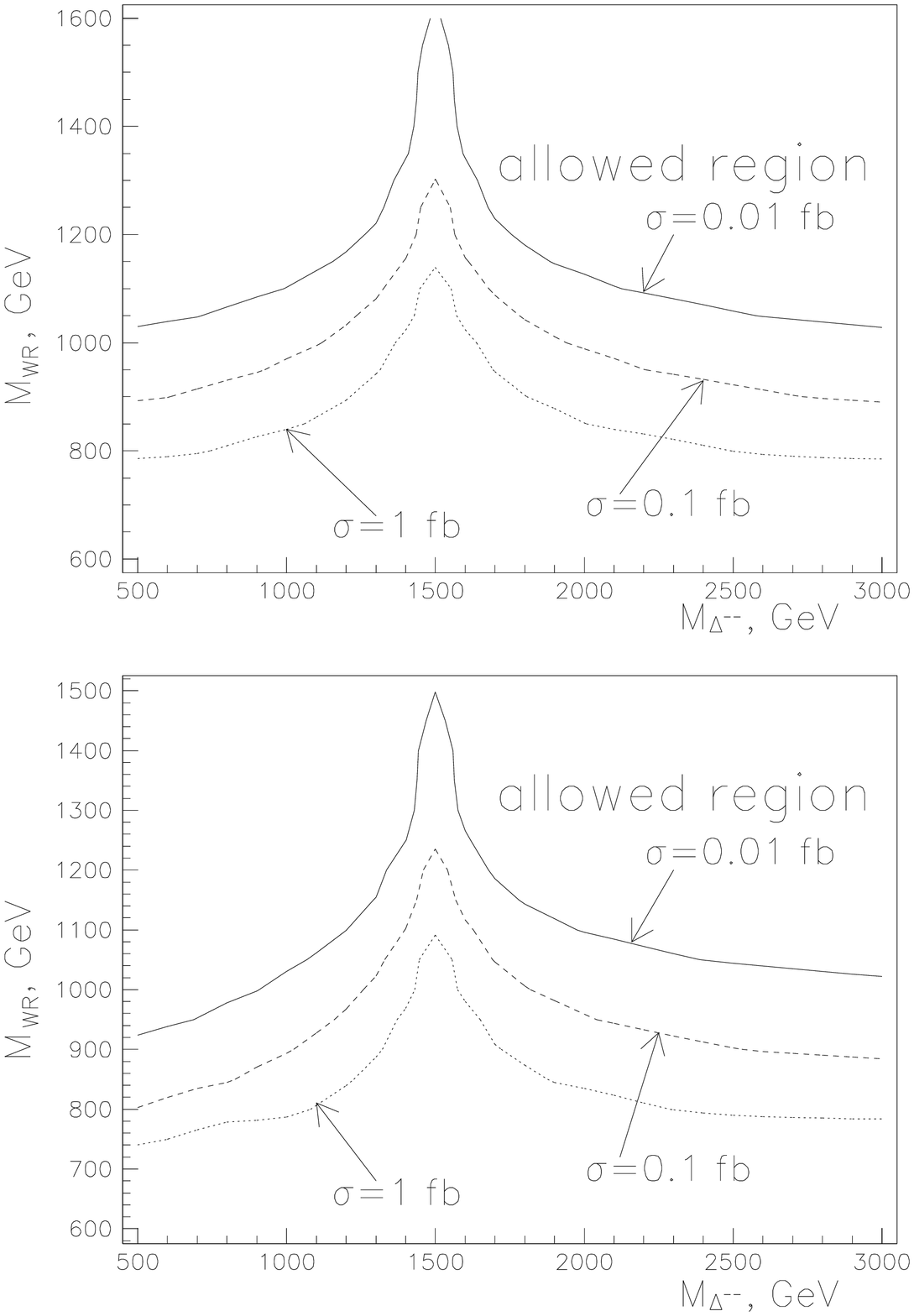}
\caption{
Contourlevels
at $\sigma=0.01 \; $ fb
(30 events per year),
 $\sigma=0.1 \; $ fb
(300 events per year), 
 $\sigma=1 \; $ fb
(3000 events per year)
for the processes with 1 $b$-jet
or with light-quarks only in the final state
(see comments in the text)  
for the energy $E=1.5 \; TeV $, 
and the right-handed neutrino masses:
$m_{\nu_2}=1.5 \; $ TeV  (on the top)
and
$m_{\nu_2}=1 \; $ TeV (in the bottom).}
\label{s15lq}
\end{figure}

\end{document}